# Laser Exfoliation of Graphene from Graphite

*Brahmanandam Javvaji*[a,b], *Ramakrishna Vasireddi*[a,c], *Xiaoying Zhuang*[b], *D Roy Mahapatra*[a*], *Timon Rabczuk*[d**]

[a]Department of Aerospace Engineering, Indian Institute of Science, Bangalore 560012, India.
[b]Chair of Computational Science and Simulation Technology,
Faculty of Mathematics and Physics, Leibniz Universität, Hannover 30167, Germany.
[c]Synchrotron SOLEIL, Saint-Aubin, Gif-sur-Yvette, 91192, France.
[d]Institute of Structural Mechanics, Bauhaus University of Weimar, 99423 Weimar, Germany.
*Corresponding author email: roymahapatra@iisc.ac.in
**Corresponding author email: timon.rabczuk@uni-weimar.de

**Abstract**

Synthesis of graphene with reduced use of chemical reagents is essential for manufacturing scale-up and to control its structure and properties. In this paper, we report on a novel chemical-free mechanism of graphene exfoliation from graphite using laser impulse. Our experimental setup consists of a graphite slab irradiated with an Nd:YAG laser of wavelength 532 nm and 10 ns pulse width. The results show the formation of graphene layers with conformational morphology from electron microscopy and Raman spectra. Based on the experimental results, we develop a simulation set up within the framework of the molecular dynamics that supplies the laser-induced electromagnetic energies to atoms in the graphite slab. We investigate the influence of different laser fluence on the exfoliation process of graphene. The variations in inter-layer interaction energy and inter-layer distance are the confirmative measures for the possible graphene layer formation. The simulation results confirm the exfoliation of a single layer graphene sheet for the laser power ranging from $100 \times 10^{-14}$ to $2000 \times 10^{-14}$ J/nm$^2$. With an increase of laser fluence from $2000 \times 10^{-14}$ to $4000 \times 10^{-14}$ J/nm$^2$, there is an increase in the graphene yield via the layer-after-layer exfoliation. The bridging bond dynamics between the successive graphene layers govern the possibility of second-layer exfoliation. The experimental and simulation observations are useful and promising for producing chemical-free graphene on a large scale for industrial and commercial applications.

## 1. Introduction

Graphene is a two-dimensional allotrope of carbon, having several applications in various fields and technologies[1], which urge to find several routes for massive high-quality production. According to published literature, the early stage of graphene synthesis studies used mechanical exfoliation from graphite using scotch tape[2]. In this method, graphene is peeled from the graphite slab. For large scale production of graphene, this method is not the right choice because it leaves residues in the exfoliated graphene leading to poor quality[3]. Chemical routes promise for the production of large scale graphene. Epitaxial growth of graphene[4,5], chemical intercalation[6–9], and chemical vapor deposition method[10–12] are well known for the large scale production with compromise on the quality. For further details about the synthesis procedures of graphene are found from the literature[13–15]. Furthermore, manufacturing problems such as long processing time, weak bonding between the graphite layers are cleaved using a solvent, high-temperature post-treatments, etc. Therefore, chemical-free routes are promising for the high-quality graphene production.

Mechanical exfoliation is a well-developed technique for large scale production of graphene, making it the current state of the art fabrication process. For example, graphene is peeled from graphite using a three-roll milling machine[16,17]. Shear exfoliation of graphene from graphite also gives large-scale quality graphene[18–20]. The basic idea of graphite exfoliation by mechanical methods is based on the anisotropic elastic stiffness coefficients of the graphite slab. Very small shear coefficients between the layers promise for cleavage of graphene under shear exfoliation. A wedge based cleavage of few-layer graphene was reproduced with molecular dynamics simulations[21]. Impacting carbon nanotubes to the solid substrate with very large velocities result in the formation of graphene[22]. Carbon nanotubes at high strain rate deformation in cryo conditions can lead to the production of graphene[23]. In addition, the recent investigation explore the promising and high yield possibility of graphene exfoliation from the application of mechanical shear forces[24]. However, these techniques require precise control of the mechanical loading to produce a good quality graphene. A recent review summarizes the modeling efforts for the growth of various 2D materials and several challenges involved[25]. Laser-induced exfoliation is another advanced technique which is aimed to get large scale and high-quality graphene[26–29].

Recent studies have shown the advantage of laser-based exfoliation and similar mechanisms of synthesis. Lasers for industry scale processing are available in the range of IR to deep UV with broad variation in the laser pulse-width (even femtoseconds) and high amount of laser beam fluences without use of an external converging lens. Ultrafast laser ablation studies give the initial understanding of laser-driven exfoliation[30]. Conventional Nd:YAG laser used to exfoliate graphene[31] and exfoliation in nitrogen as solvent has been studied[32]. To speed up the chemical vapor deposition and intercalation methods, a laser is used as an external energy source[15]. A recent study reports the selective synthesis of graphene or graphene oxide quantum dots by controlling the wavelength of pulsed laser irradiating on the multi-wall carbon nanotubes[33]. An *ab* initio method has been developed to produce graphene from laser irradiation using computer simulations[34]. In that report, time-dependent density functional calculations are used for dealing with electrons and the molecular dynamics simulations are carried for ion dynamics. The laser electromagnetic field irradiation is assumed as uniform throughout the periodic 2x2 supercell, which leads to a uniform separation of the graphene layer from the graphite slab. In experimental situations, laser irradiation has been focused on a portion of the graphite slab, which induces a local heating and upon the layer exfoliation can take place by the thermal conduction. However, the simulation-based study gives a mechanistic understanding that the laser-induced forces overcome the inter-layer interactions and the graphene exfoliation is possible. An equivalent approach is to model the dynamic behavior based on molecular dynamic force field derived from inter-atomic potential where the effects of charge and equivalent effect of the electromagnetic field can be incorporated. The incorporation of electromagnetic fields of the laser into the classical molecular dynamics force fields is successful in simulating the molecular trajectories under strong laser fields[35], efficient machining of silicon carbide into different nanostructures[36], generating nanoclusters of aluminum by ultrafast laser ablation method[37], etc.

In this work, we considered a graphite slab irradiated with laser. The laser-induced forces are estimated from the fundamental behavior of charged particles in electromagnetic fields. A molecular dynamics model is setup using the charge based potential and the electromagnetic fields are subjected as a force to the atoms. Various quantities like inter-layer energy, inter-layer distance, bridging bond dynamics are utilized to understand the exfoliation process. An experimental setup is also developed for the initial observation of graphene exfoliation.

## 2. Experimental Scheme and Simulation Method

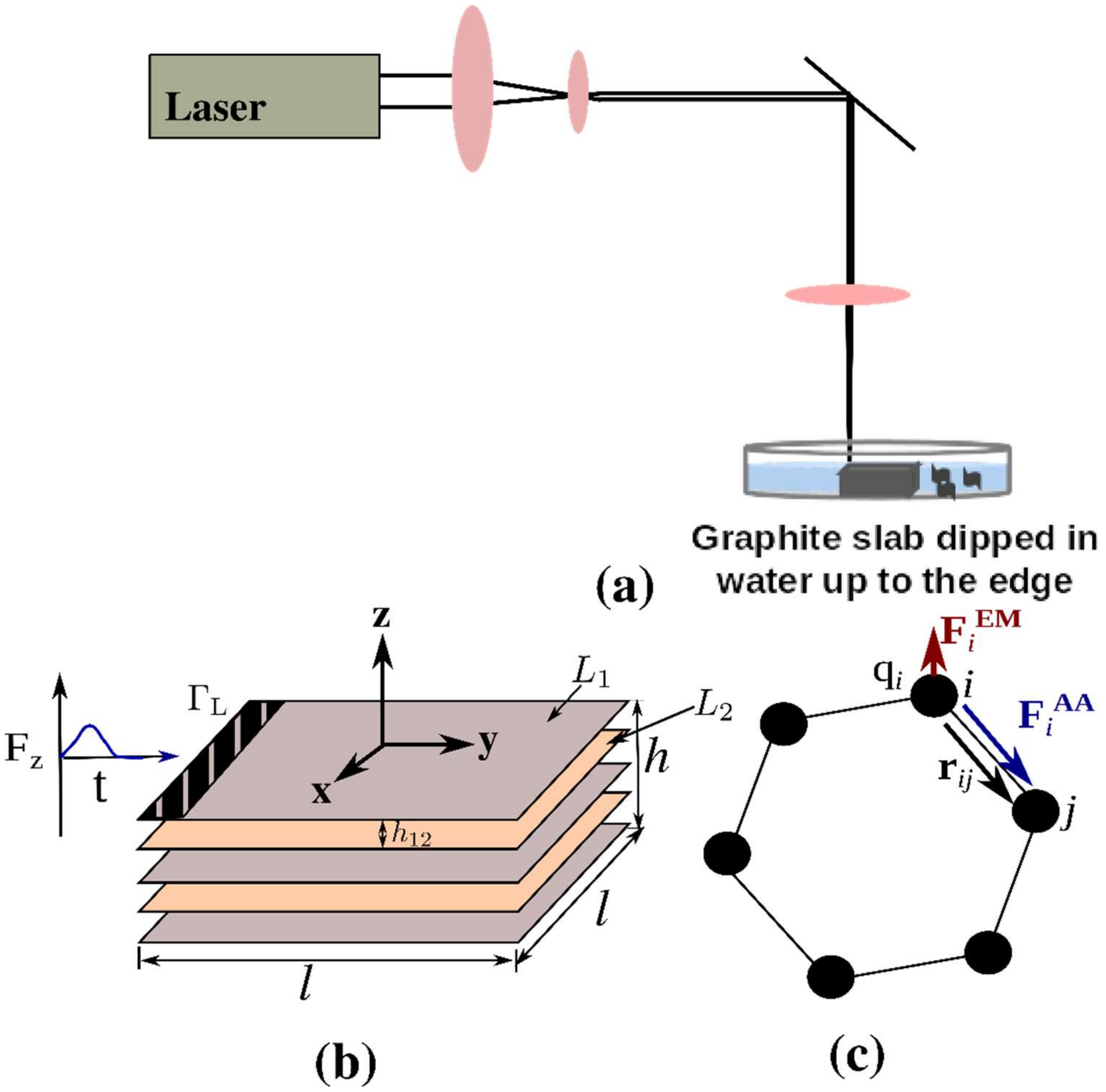

**Figure 1: (a) Illustration of the experimental setup for graphene exfoliation from graphite slab with a laser. (b) Schematic view of graphite slab illuminated with laser illuminating on the marked area ($\Gamma_L$)and (c) atomic representation of carbon ring subjected to different forces.**

**Figure 1(a)** explains the experimental setup, the schematic of the graphite slab in the water bath subjected to laser. Q-switched Nd:YAG laser with a wavelength of 532 nm, 10 ns duration, 10 Hz repetition rate is used in the experiment. Graphite slab target of 5 mm thick and 1 cm diameter was submerged in a water bath such that the top-left edge of the surface of graphite is exposed. The laser beam was focused through a convex lens to nearly 1 $mm^2$ spot size on the graphite target. The effects of laser energy range from 105 to 370 mJ on graphene layer formation as an end product were studied. After exposure to the laser beam, the water in the bath was evaporated at room temperature and the residue was collected for further characterization. For further understanding of the mechanism of graphene formation, we develop a simulation set up within the framework of the molecular dynamics.

A square-shaped graphite slab of length $(l)$ 5 nm and thickness $(h)$ 2 nm is considered. Seven graphene layers were stacked with ABA sequence, where A and B corresponds to the repetitive atomic arrangement in the first layer $L_1$and second layer $L_2$, respectively (as shown in **Figure 1(b)**. The initial distance of separation between $L_1$ and $L_2$ is represented by $h_{12}$, which is equal to $0.335$ nm. The schematic of the graphite slab with dimensions and other physical variables is shown in **Figure 1(b)**. A zoomed portion of **Figure 1(b)**shows the hexagonal arrangement of the layer $L_1$ (see **Figure 1(c)**). The atomic arrangement has the internal interaction energy which is estimated from the many-body inter-atomic potentials.

As indicated in **Figure 1(c)**, the atoms are subjected to forces like atom-to-atom interaction force $\mathbf{F}_i^{\mathrm{AA}}$ on atom $i$ and forces due to electromagnetic (laser) irradiation $\mathbf{F}_i^{\mathrm{EM}}$. The total force $\mathbf{F}_i$ on the atom $i$ is

$$\mathbf{F}_i = \mathbf{F}_i^{\mathrm{AA}} + \mathbf{F}_i^{\mathrm{EM}}. \tag{1}$$

Force $\mathbf{F}_i^{\mathrm{AA}}$on atom $i$ is calculated using the first-order spatial derivative of the inter-atomic potential $(U)$defined over a system of atoms

$$\mathbf{F}_i^{\mathrm{AA}} = -\frac{\partial U}{\partial \mathbf{r}_i}, \tag{2}$$

where $\mathbf{r}_i$ is the position vector of atom $i$. We have employed Charge Optimized Many-Body (COMB) potential to determine the atomic configuration as well as point charges associated with each atom. For further details about the COMB potential, refer to the original literature[38]. Using COMB potential, the atomic interactions within the layer of graphite (graphene) are correctly reproduced[39]. The dynamic charge-dependent interactions accurately capture the interfacial interactions between hydrocarbons and solid surfaces[39,40]. The graphite properties like cohesive energy, inter-layer spacing, bond length, and density computed using COMB potential are in god agreement with experimental measurements[41].

Force $\mathbf{F}_i^{\text{EM}}$ on atom$i$ is approximated to the Lorentz force acting on the atom having charge $q_i$moving with velocity $\mathbf{v}_i$ in a background electromagnetic field described by electric field $\mathbf{E}(\mathbf{r}_i, t)$ and magnetic field $\mathbf{B}(\mathbf{r}_i, t)$, that is,

$$\mathbf{F}_i^{\text{EM}} = q_i\mathbf{E}(\mathbf{r}_i, t) + q_i\big(\mathbf{v}_i \times \mathbf{B}(\mathbf{r}_i, t)\big) \tag{3}$$

where $\mathbf{E}(\mathbf{r}_i, t)$ and $\mathbf{B}(\mathbf{r}_i, t)$ are primarily due to lasing and secondarily due to the atomic charge distribution that evolves the long-range dynamics. Laser properties are introduced in terms of laser power $L_{\text{p}}$, laser pulse-width $\tau$ and laser wavelength $\lambda$. The laser beam is incident on the area $\Gamma_{\text{L}}$ (shaded region in **Figure 1(b)**). The electric and magnetic field strength for the laser beam are estimated from the laser intensity $I_{\text{L}}$ which is the power flow per unit area[42], also known as pointing vector $(\mathbf{S})$,

$$I_{\text{L}} = \|\mathbf{S}\| = \frac{1}{\mu}|\mathbf{E}_0 \times \mathbf{B}_0| \tag{4}$$

where $\mathbf{E}_0$ and $\mathbf{B}_0$ are the amplitudes of fields in an electromagnetic wave, $\mu$ is the magnetic permittivity of the graphite slab. Relationship between $\mathbf{B}_0$ and $\mathbf{E}_0$ can be obtained from Maxwell's equation as

$$\nabla \times \mathbf{E} = -\frac{d\mathbf{B}}{dt} \tag{5}$$

and assume that electric and magnetic fields of a monochromatic wave have identical phase $\big(exp(ik\hat{\mathbf{k}} \cdot \mathbf{r} - i\omega t)\big)$ with amplitudes $\mathbf{E}_0$ and $\mathbf{B}_0$, where $k$ and $\omega$ are the wavenumber and frequency of the laser. $\hat{\mathbf{k}}$ is the unit vector representing the incident wave direction.

Substituting the expressions for the fields into Eq. (5) leads to the incident magnetic field, which is

$$\mathbf{B}_0 = \frac{k}{\omega}\left(\hat{\mathbf{k}} \times \mathbf{E}_0\right). \tag{6}$$

Putting back Eq. (6) back to Eq. (4), $E_0$ can be estimated as

$$E_0 = \sqrt{\frac{I_L \mu \omega}{k}}\,. \tag{7}$$

By knowing the electric field amplitude $E_0$ from Eq. (7), we design the time variation of **E** to achieve the pulse width $\tau$, by involving the modulation frequency $\nu_m$, which is

$$\mathbf{E}(t) = E_0 \zeta \left(t - \frac{1}{\nu_m}\right) \sin(2\pi\nu t)(1 - \cos(2\pi\nu_m t)\,)\,\hat{z} \tag{8}$$

where $\zeta$ is a unit step function that defines the number of waves over a pulse-width. $\zeta$ is zero for $t - 1/\nu_m$ is less than 0. The time variation of **B** is identical to Eq. (6). The time-dependent functions of **E** and **B** are substituted in Eq. (3) for laser-induced forces. Note that, the given **E** and **B** fields in our simulation problem are assumed as uniform in the illuminating region ($\Gamma_L$). Since this region is several orders smaller compared to the experimental zone of irradiation, the intensity at the central portion of the Gaussian beam profile is approximated as uniform over the given region $\Gamma_L$. This approximation can be relaxed further by considering the graphite system with hundreds of nanometer size. However, such systems are computationally expensive in finding the dynamical charges, which are involved in the effective laser force fields. Moreover, the application of time variation of electromagnetic fields in Eq. (8) and associated laser force fields to the atoms is certainly a promising approach to model the laser treatment. Before applying laser excitation, the graphite slab is thermally equilibrated to a temperature of 300 K using Nosé-Hoover thermostat[43,44] for a period of 10 ps with a time step of 1 fs. The bottom of the graphite slab is restrained. Atoms in the excitation region ($\Gamma_{\mathrm{I}}$) are subjected to the electromagnetic forces ($\mathbf{F}_i^{\mathrm{EM}}$) after thermal equilibrium has been attained. We relax the thermostat condition after the equilibration period, which is to capture the laser-induced heating in the graphite slab. At each 0.1 ps simulation time step, we determine the atomic configuration along with various computed atomic properties. We analyze the exfoliation process using the thermal distribution and the distance of separation between the layers over time. All simulations were performed using the open-source MD simulation code large-scale atomic/molecular massively parallel simulator (LAMMPS)[45].

## 3. **Results and Discussions**

To investigate the microstructures and surface morphology of the as-synthesized graphene materials, field emission scanning electron microscopy (FESEM) and Raman spectrometry experiments were performed. The graphene layer formation as a function of laser fluence is shown in **Figure 2**. **Figure 2(a)** shows that for laser fluence at $30 \times 10^{-14}$ J/nm$^2$, no exfoliation was obtained from the substrate, which indicates that the laser energy is not sufficient to peel the graphene layer. The graphene layer separation has started from $63.5 \times 10^{-14}$ J/nm$^2$, as shown in **Figure 2(b)**. Thin graphene layers are observed at higher laser fluence $104 \times 10^{-14}$ J/nm$^2$ (see **Figure 2(c)**) and corresponding Raman spectra are shown in **Figure 2(d)**. The observation of multi-layer graphene at this laser fluence range is in better comparison with earlier experimental reports[31,46].

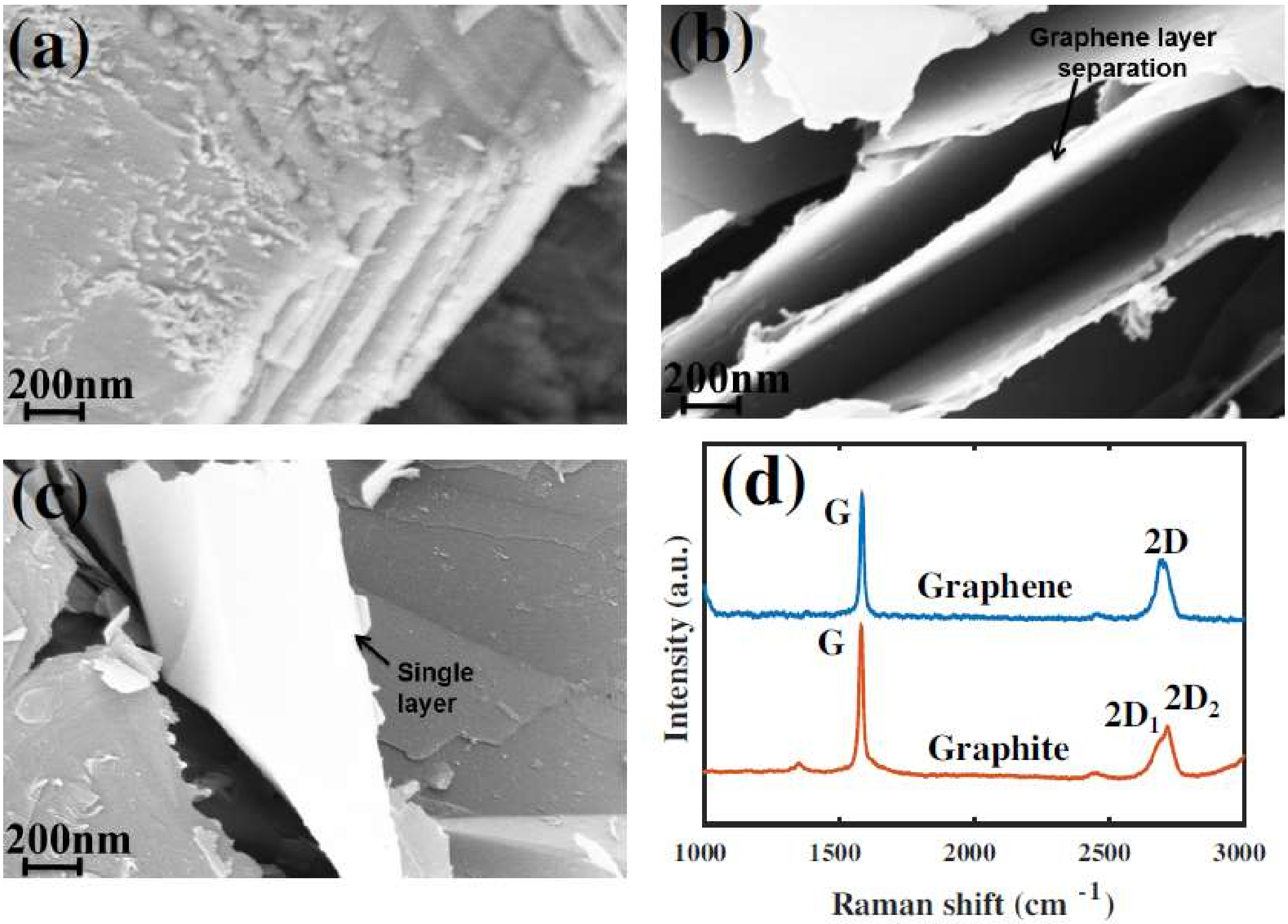

**Figure 2: FESEM images of graphene layers obtained at (a) $30 \times 10^{-14}$ J/nm², (b) $63.5 \times 10^{-14}$ J/nm² and (c) $105 \times 10^{-14}$ J/nm². (d) Raman spectra at 514 nm for graphite slab and exfoliated graphene layers.**

The 2D peak position is found at 2692 $cm^{-1}$, which is in agreement with earlier reports[47]. From the literature, the ratio of intensity of 2D peak to G peak is two. The peak frequency is matching with the reports, whereas it differed with the intensity. The intensity of the 2D peak represents the number of layers in graphene. If a multi-layer graphene is considered, the intensity of the 2D peak is reduced. This represents that, from our experiment, we achieved a multi-layer graphene. Fine-tuned control on the laser fluence and laser pulse width can achieve single-layer to multi-layer graphene sheets. To verify the controllability on the graphene exfoliation and to have a better understanding of the exfoliation process, we proceed to perform simulations. Note that the power of irradiation used in the experiments is quite high but the beam diameter is large (about 9.5 mm), and it may require further focusing of the beam to achieve controllable energy density for exfoliating a single graphene layer. There exists a strong possibility to exfoliate a single graphene sheet with lower laser powers and pulse width. The employed laser properties in simulations are tabulated in **Table 1**. We have varied the laser power from 0.2 to 20 W irradiating on the area $\Gamma_I$ (5 nm$^2$) to achieve a comparable laser fluency used in experiments. The calculated laser fluence per each simulation step is ranging from where $40 \times 10^{-14}$ to $4000 \times 10^{-14}$ J/nm$^2$ the frequency is equal to the experimental laser frequency. Note that, the value of laser fluence used in experiments are according to the actual Gaussian beam and not on the localized area of irradiation. Whereas, in simulations, the computation of laser fluence is according to the selected illumination region.

**Table 1: Experimental and simulation conditions**

| Quantity | Experiment | Simulation |
|---|---|---|
| Frequency | 563 THz | 600 THz |
| Wavelength | 532 nm | 500 nm |
| Laser spot size | 1 mm$^2$ | 5 nm$^2$ |
| Laser pulse width | 10 ns | 10 ps |
| Maximum laser power | $36 \times 10^6$W | 20 W |
| laser fluence range | $30 \times 10^{-14}$ to $105 \times 10^{-14}$ J/nm$^2$ | $40 \times 10^{-14}$ to $4000 \times 10^{-1}$ J/nm$^2$ |
| Graphite slab | $0.01 \times 0.01 \times 0.005$ m$^3$ | $5 \times 5 \times 2$ nm$^3$ |

We analyze the obtained simulation data with the help of inter-layer energy and the inter-layer distance. The time variation of these quantities indicates the chance of exfoliation. The inter-layer energy is the interaction energy among atoms between successive layers in a graphite slab. The average energy of atoms in layer $L_1$ is represented as $U(L_1)$ and average energy of atoms in layer 2 is $U(L_2)$. The difference between $U(L_1)$ and $U(L_2)$ gives the inter-layer energy between layers 1 and 2. The difference between average z-coordinate of atoms in layer 1 and layer 2 gives the distance of separation $h_{12}$ between them. **Figure 3(a)** and **(b)** show the time variation of quantities $U(L_2) - U(L_1)$ and $h_{12}$, respectively.

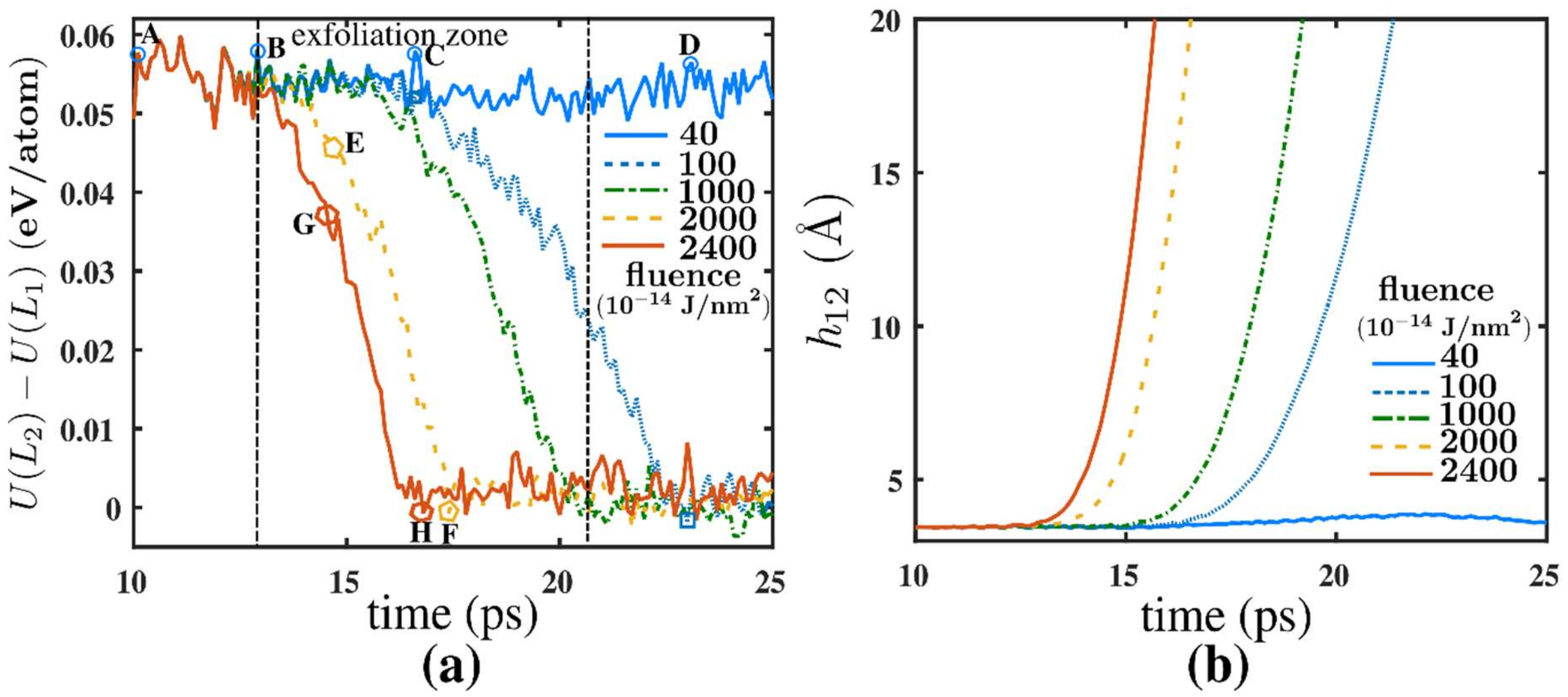

**Figure 3: The variation of (a) inter-layer energy $U(L_2) - U(L_1)$ and (b) inter-layer distance $h_{12}$ as a function of time. The different line styles indicate the time response at different laser fluences. Markers A to H indicate the selected time instances for monitoring the atomic configurations.**

For laser fluence $40 \times 10^{-14}$ J/nm$^2$, the inter-layer energy is equal to 0.05 eV/atom with an inter-layer separation of 3.315 Å, which indicates the absence of graphene exfoliation at a laser fluence of $40 \times 10^{-14}$ J/nm$^2$. The timely evolution of the atomic system further confirms this observation. **Figure 4(a)** to **(d)** correspond to the atomic configuration at selected time instances A, B, C and D in **Figure 3(a)** when laser fluence is $40 \times 10^{-14}$ J/nm$^2$. The laser pulse hits the graphite slab between 10 to 20 ps (according to the pulse width of the laser). Blue color indicates the portion of the top graphene layer used for the laser impulse and the remaining portion of the top layer $L_1$ marked with black color. Orange color indicates the next-to-top layer $L_2$ sub-surface layer. **Figure 4(a)** represents the equilibrium atomic configuration. From **Figure 4(b)**, there is a slight lift-off in the irradiated region because of the laser impulse. At 17.5 ps (point C), **Figure 4(c)** shows a noticeable lift in the loading region. At 22 ps (point D), when the laser is turned off, the atomic configuration settles down to the equilibrium state. Overall, $40 \times 10^{-14}$ J/nm$^2$ laser fluence is not sufficient to exfoliate the graphene. This observation is in close agreement with our experimental result of no exfoliation when laser fluence is about $30 \times 10^{-14}$ J/nm$^2$, which supports that current numerical modeling and simulation settings are accurately capturing the real process of exfoliation.

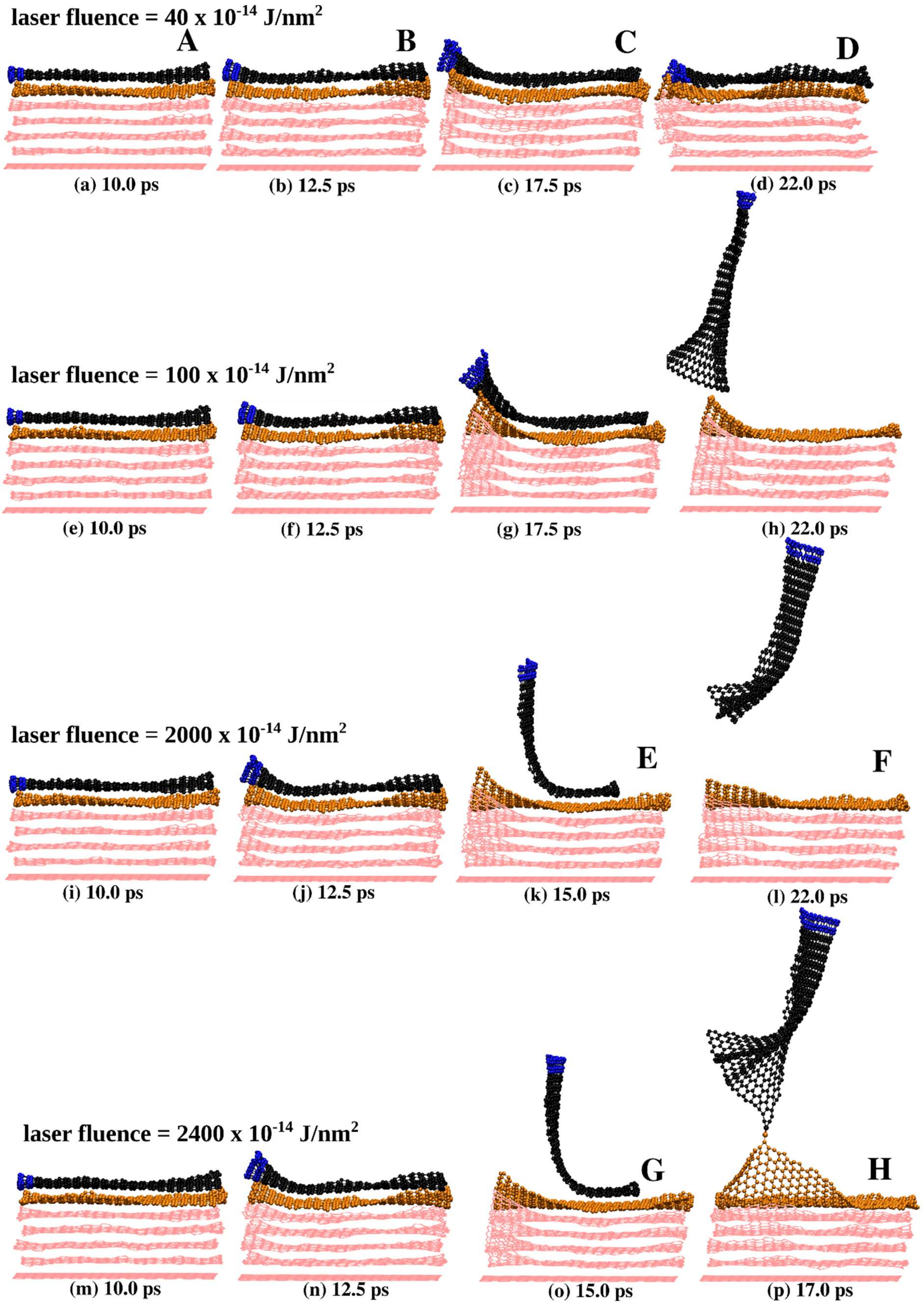

**Figure 4:Laser-induced atomic evolution of graphite slab at various laser powers and at different time instances. Blue color indicates the portion of the top graphene layer used for the laser impulse and the remaining portion of the top layer marked with blue color. Orange color indicates the next-to-top layer. The remaining layers are colored with pink.**

In the case of $100 \times 10^{-14}$ J/nm$^2$ laser fluence, we identify a decrease in the $U(L_2) - U(L_1)$ between 17.5 and 22.0 ps times (From **Figure 3(a)**). **Figure 4(g)** shows that the separation of $L_1$ from $L_2$ gets initiated at 17.5 ps. Since a set of atoms are moving away from the interaction neighborhood of $L_2$, the energy shows a reduced value. The separation process continued further and, it completely separates the graphene sheet $L_1$ (**Figure 4(h)**). At 22 ps time, the interlayer binding energy tends to zero. The interlayer distance starts increasing and shoots to a high value sometime between 17.5 ps and 22.0ps time. The drops in the binding energy, variation of interlayer distance and, atomic configurations confirm about the exfoliation of graphene at laser fluence $100 \times 10^{-14}$ J/nm$^2$. This observation is also in accordance with our experimental result of graphene exfoliation at $63.5$ to $105 \times 10^{-14}$ J/nm$^2$.

The current optical setup in experiments reduces the laser spot size to 1 mm and produces the laser fluence of $40$ to $105 \times 10^{-14}$ J/nm$^2$. In this laser fluence regime, the experiments and simulations results have very close agreement. We proceed to increase the laser fluence further (up to $4000 \times 10^{-14}$ J/nm$^2$) in simulations, to study the effect on the exfoliation process. However, further increment of laser fluence in experiments require proper focusing, which could be challenging because the graphite slab has to be aligned and cannot move away if a water bath was used.

The variation of binding energy and interlayer distance is given at laser fluence $1000 \times 10^{-14}$ J/nm$^2$. There is an increase in the slope of the interlayer distance variation with the increase in laser fluence from $100$ to $1000 \times 10^{-1}$ J/nm$^2$, which indicates the speed-up in the process of exfoliation. For $2000 \times 10^{-14}$ J/nm$^2$ laser fluence, the separation is initiated at 12.5 ps time. The inter-layer energy tends to zero between 12.5 ps and 17.5 ps. In the variation of inter-layer energy at $2000 \times 10^{-14}$ J/nm$^2$ laser fluence, two different time instances are selected (E and F). Nearly half of the graphene sheet is cleaved from the slab (see **Figure 4(k)**). At 17.5 ps, there is a complete separation of graphene and the inter-layer binding energy reaches zero. The response at laser fluence 240 J/cm$^2$ shows a little speed up in the exfoliation time. At time 15ps, the binding energy level is decreased compared to $2000 \times 10^{-14}$ J/nm$^2$ (point G), and more than half of the graphene sheet is exfoliated (**Figure 4(o)**). At 17.0ps, graphene sheet $L_1$ is completely exfoliated from graphene slab (from **Figure 4(p)**).

The simulation indicates graphene exfoliation at laser fluences greater than or equal to $100 \times 10^{-14}$ J/nm$^2$, which is confirmed by the calculations of interlayer binding energy and distance of separation. Because of the electromagnetic force field, these layers start moving away from the rest of the slab. With the increase in force, the displacement and velocity of the atoms are also increasing. We have used the displacement or the distance of separation to explain the exfoliation process. The remaining quantity like the velocity of atoms is also utilized to analyze the exfoliation process in terms of temperature. To analyze the spatial and temporal distribution of temperature in $L_1$, the atoms in that layer are divided into several rectangular bins along the y-direction and the temperature of these bins was computed at each time step. **Figure 5** shows the temporal and spatial distribution of the temperature of the layer $L_1$ for different laser fluence $40, 100$ and $2000 \times 10^{-14}$ J/nm$^2$.

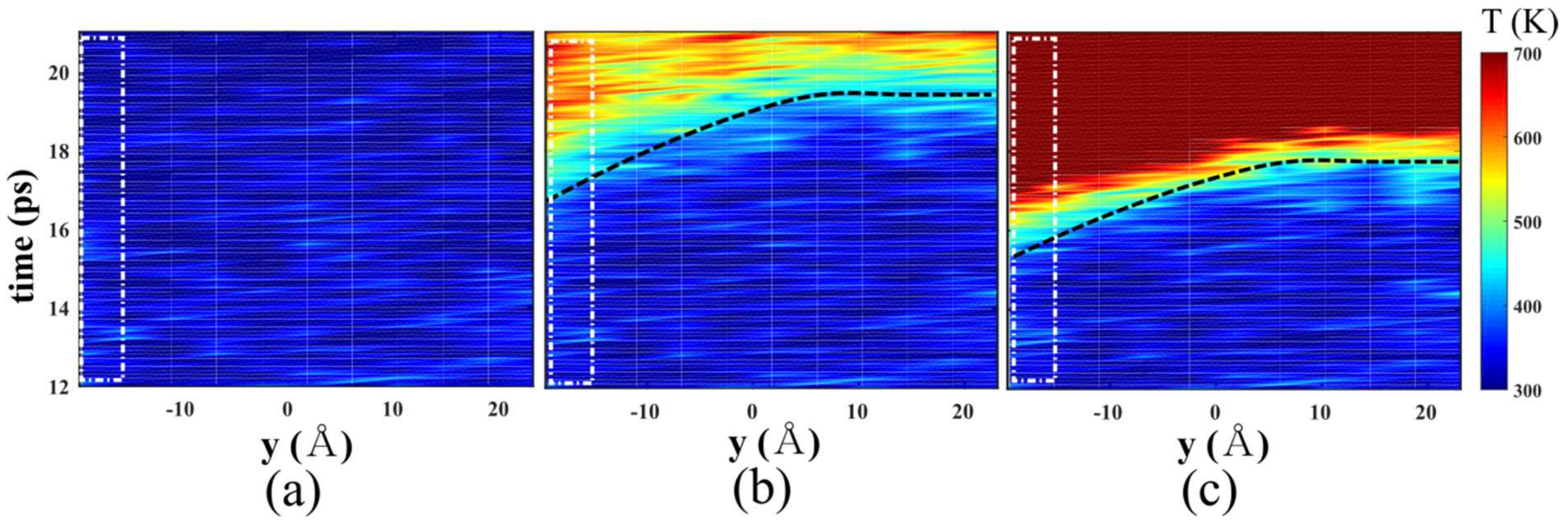

**Figure 5: Spatial and temporal evolution of the temperature of the layer $L_1$ at laser fluence (a) $40 \times 10^{-14}$ J/nm², (b) $100 \times 10^{-14}$ J/nm² and (c) $2000 \times 10^{-14}$ J/nm² . The rectangle represents the illuminating region $\Gamma_I$. The dashed line indicates the spatial conduction of thermal energy. $y$coordinate of atomic system is considered for monitoring the spatial variation of temperature, according to Figure 1(b). The color bar shows the temperature scale.**

The laser-induced force increases the kinetic energy of atoms in $L_1$. With the increased kinetic energy, the atoms in $L_1$ move away from the graphite slab. The kinetic energy gain is related to the laser-induced force, which is acting on the atoms. Note that in the simulation we are not controlling the temperature after the thermal equilibration, which is to capture the laser-induced heating. The rectangle region marked in **Figure 5** represents the irradiation region $\Gamma_I$. The dashed line indicates heat conduction induced thermal energy gradient. The color bar shows the scale of variation in temperature. The maximum temperature reaches a maximum of 700 K. In the case of $40 \times 10^{-14}$ J/nm² laser fluence, the temperature varies between 300 to 400 K (**Figure 5(a)**), where the kinetic energy gain is small and is not sufficient to exfoliate the graphene. For $100 \times 10^{-14}$ J/nm² laser fluence, the variation of temperature with time in the loading region is about 300 to 700 K (**Figure 5(b))**. The rise in temperature leads to stronger debonding of the illuminating region from the graphite slab, which represents that this region obtained sufficient kinetic energy for lift-off or increasing the $h_{12}$ in response to the laser impulse. Also, an energy transfer from the loading region to the neighboring atomic regions exists via the heat conduction process. The heated neighboring region starts increasing the $h_{12}$ further, which makes $L_1$ moving away from the slab and leads to complete separation. When laser fluence is $2000 \times 10^{-14}$ J/nm$^2$, the observed exfoliation process is complete within 2 to 5 ps time, where the laser irradiation is still active. Because of the short exfoliation time, the exfoliated graphene layer attains very large kinetic energy, which is represented as dark red color in **Figure 5(c)**.

Another observation from the atomic snapshots is that there is a second layer lift-off identified from **Figure 4(p)** at laser fluence $2400 \times 10^{-14}$ J/nm$^2$. It also observed that there develops a strong bridging bond between edge atoms of layer $L_1$ and layer $L_2$. Because of this bridging bond, the laser-induced heat and forces get transferred from $L_1$and $L_2$, which helps to lift the layer $L_2$.The increased energy transfer avoids the interaction of $L_2$ with layers below it and finally leads to complete exfoliation of $L_2$. We evaluate the bond energy and bond length for this bridging bond and compare it with the bond between carbon atoms in single-layer graphene.

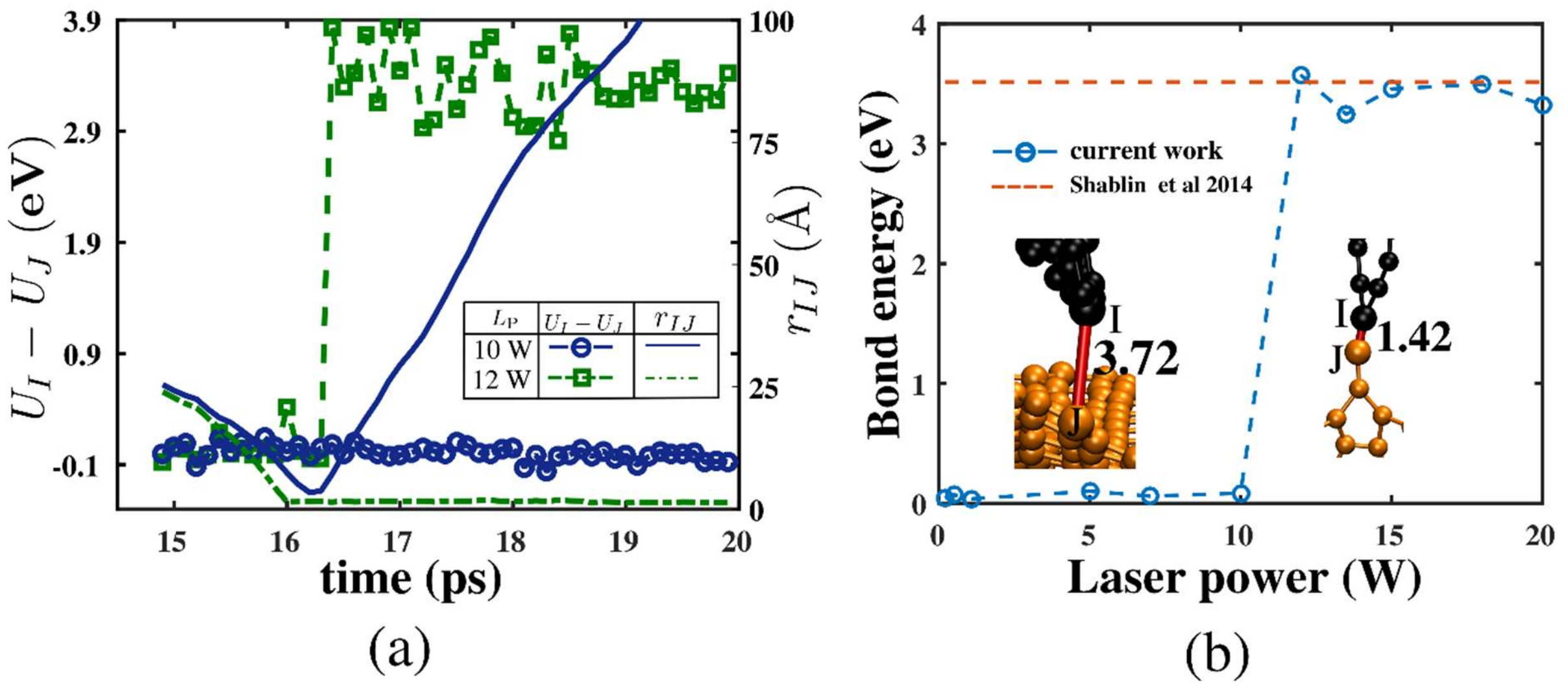

**Figure 6:Time history of inter-atomic energy $U_I - U_J$ between bridging atoms and the inter-atomic distance $r_{IJ}$ at laser fluence 2000 and $2400 \times 10^{-14}$ J/nm$^2$. (b) Variation of bridging bond energy at different laser fluences. Solid line corresponds to the value from Shablin et al.**[48]

**Figure 6(a)** shows the time response of inter-atomic energy difference ($U_I - U_J$) between atoms $I$ and $J$ and their inter-atomic distance $r_{IJ}$. Note that atoms $I$ and $J$ belong to the layers $L_1$ and $L_2$, respectively. The inset of **Figure 6(b)** shows the zoomed atomic configurations near the bridging bond for $2000 \times 10^{-1}$ J/nm$^2$ and $2400 \times 10^{-14}$ J/nm$^2$ laser fluence. The initial atomic energy between atoms $I$ and $J$ is nearly equal to -0.01 eV when laser fluence is $2000 \times 10^{-14}$ J/nm$^2$, which represents that these atoms are having only a weak long-range interaction (Van Der Waals and Coulombic interactions). The atomic distance between them decreases from 25 Å to 3.72 Åbetween 15 to 16.3 ps. After 16.3 ps, the inter-atomic distance gradually increases to very high values, which indicates that atom $I$ from $L_1$ move very far away from atom $J$, which belongs to $L_2$.The time variation of $U_I - U_J$and $r_{IJ}$ supports the absence of second layer exfoliation when the laser fluence is $2000 \times 10^{-14}$ J/nm$^2$. In the case of $2400 \times 10^{-14}$ J/nm$^2$, the inter-atomic energy is equal to -0.01 eV between 15 to 16.3 ps. At 16 ps, the energy shows a small rise, which represents an increased attraction between these two atoms. Such atomic attraction leads to a decrease in the distance between atoms from 25 Å to 1.5 Å. Further, the developed atomic attraction transforms into a strong interaction at 16.3 ps and increase$U_I - U_J$. After 16.3 ps, the bond distance nearly concentrates to 1.42 Å and the bond energy oscillates near 3.5 eV. The bridging bond energy and length variation with time for laser fluence 240 J/cm$^2$ indicate that the layers $L_1$ and $L_2$ are strongly bonded with each other and helps to exfoliate the second layer.

We have further increased the laser fluence from 2400 to $4000 \times 10^{-14}$ J/nm$^2$, to verify the existence of the bridging bond. **Figure 6(b)** shows the variation of inter-atomic energy (bridging bond energy) with different laser powers. Between 40 to $2000 \times 10^{-14}$ J/nm$^2$, the energy of the bridging bond is nearly zero, which represents that the attraction between the atoms across the layers is not sufficient to form a bond. For laser fluence greater than $2000 \times 10^{-1}$ J/nm$^2$, the bond energy is about 3.5 eV. This value is in agreement with the bond energy between carbon atoms in *sp*$^2$ hybridization[48]. The bridging bond helps in transferring the laser energy (kinetic energy) to the successive layer. $L_1$ Layer 1 starts pulling $L_2$ layer 2 with the help of a bridging bond and a complete exfoliation of the second layer is observed. After the second layer is exfoliated from the graphite slab, the bridging bond is not stable. The system relaxes to a stable configuration by developing a weak interatomic interaction between the layers.

## 4. **Conclusions**

In this study, we performed both simulations and experiments to exfoliate the graphene from a graphite slab. Spectroscopic results indicate the existence of several layered graphene sheets. We set up a simulation scheme within the framework of the molecular dynamics to replicate the laser-induced irradiation process. We model the laser irradiation-induced force on the atoms. With these forces, atoms also developed higher temperature by gaining kinetic energy. The observation of no graphene exfoliation at lower fluence (nearly $40 \times 10^{-14}$ J/nm$^2$) is identical in both simulation and experiments. The variation of interatomic energy between layers and interatomic distance with simulation time shows the initial observation of graphene layer exfoliation. As the laser fluence increases, the rate of change in the interatomic energy also increases. The laser irradiation-induced force increases the kinetic energy, which is observed by monitoring the temperature. Atomic snapshots indicate a bridging bond between layers at laser fluence $2400 \times 10^{-1}$ J/nm$^2$. The dynamics indicate that carbon-carbon bridging bonds exist between the layers with which the energy transfer to the second layer and beyond is possible. As a result, the second layer of exfoliation is observed clearly. Better control over the focused irradiation zone and pulse width optimization appears promising toward graphene production via chemical-free laser exfoliation.

Acknowledgment: DRM, BJ, RV thankfully acknowledge financial support from the Aeronautics Research & Development Board through the project ACECOST-Phase III at the Indian Institute of Science. DRM thankfully acknowledges DRDO Chair Professorship at the Indian Institute of Science during finalizing this paper. Authors thankfully acknowledge the use of Pulsed Laser Facility at the Department of Aerospace Engineering, and the characterization facility at the Centre for Nano Science and Engineering, Indian Institute of Science for the experiments performed. BJ also acknowledges financial support as Mari Curie Visiting Researcher at Bauhaus University, Germany through the project Multiscale Fracture.